\documentclass[preprintnumbers,showpacs,showkeys,amsmath,amssymb]{revtex4}
\usepackage{graphicx}
\usepackage{dcolumn}
\usepackage{bm}
\usepackage{amsfonts}
\usepackage{amssymb}
\usepackage{bbm}
\usepackage{bbold}
\usepackage{color}
\usepackage[export]{adjustbox}
\begin{document}
    \title{Effective Dynamics of a Conditioned Generalized Linear Glauber Model} 
    \author{Sara Kaviani}
    \author{Farhad H. Jafarpour}
    \email{farhad@ipm.ac.ir}
    \affiliation{Physics Department, Bu-Ali Sina University, 65174-4161 Hamedan, Iran}
    \date{\today}
\begin{abstract}
In order to study the stochastic Markov processes conditioned on a specific value of a time-integrated observable, 
the concept of ensembles of trajectories is recently used extensively. In this paper we consider a generic  
reaction-diffusion process consisting of classical particles with nearest-neighbor interactions on a one-dimensional lattice with 
periodic boundary conditions. By introducing a time-integrated current as a physical observable, we have found 
certain constraints on the microscopic transition rates of the process under which the effective process contains local 
interactions; however, with rescaled transition rates comparing to the original process. A generalization of the 
linear Glauber model is then introduced and studied in detail as an example. Associated effective dynamics of this model 
is investigated and constants of motion are obtained.  
\end{abstract}
\pacs{05.40.-a,05.70.Ln,05.50.+q}
\keywords{non-equilibrium systems, stochastic dynamics, driven dynamics, current fluctuations, large deviations}
\maketitle
\section{\label{sec1}Introduction}
It is known that statistical mechanics provides us with a toolbox to study rare-events
or fluctuations of time-integrated observables~\cite{AF01,PF,WWV05,S}. In order to probe the fluctuations of a dynamical observable, 
one can model the physical system by a stochastic Markov jump process. The time-average of such observable 
far from its typical value can then be studied using the concept of biased ensembles of trajectories associated with large deviations of 
that observable in equilibrium or non-equilibrium systems. Focusing on ensembles of trajectories, instead of ensembles of configurations, 
allows us to investigate dynamical behavior of the system. This can actually be achieved by considering the Ruelle's thermodynamic 
formalism for dynamical systems~\cite{Ru01,LRW}. This is analogous to the thermodynamic ensemble method for configurations 
or micro-states of standard equilibrium statistical mechanics~\cite{LP,DC,RKP}. 

In order to study the fluctuations of a dynamical observable one has to take into account those trajectories
along which the value of the observable is atypical up to an observation time which is assumed to be large. Hence, we are 
dealing with a type of conditioning on the generator of a stochastic Markovian system to have a specific value of the dynamical 
time-integrated observable over the dynamical trajectories. As a result, there is a need for a sort of weighting procedure. This 
weighting procedure is analogous to the canonical ensemble in equilibrium statistical mechanics which is just a modification 
of micro-canonical ensemble using Boltzmann weight.

From the non-equilibrium point of view, a micro-canonical path ensemble consists of specific trajectories with certain values
of an observable along each trajectory. In a canonical path ensemble, by introducing a conjugated quantity to the observable,
sometimes called the $s$ field, we can restrict the average value of the observable to the desired value. In this $s$-ensemble
of trajectories the evolution generator of the corresponding dynamics is called the tilted (biased) generator for which the
conservation of probability is violated~\cite{HJG,BTG,GJLPDW}. 

Now the question is whether the dynamics of the system which results in a specific value of a fluctuation in the steady state
can be described by a non-conditioned stochastic Markov process. The answer is positive and it has been shown that it is
possible to construct a non-conditioned Markov process called the driven (effective) process~\cite{JS,TJ}. In a long-time limit
and during a Time Translational Invariant (TTI) regime, the atypical value of the observable in the unconditioned process is equal to 
the typical value of the observable in the steady-state of the driven stochastic process~\cite {CT1, E04}. In other words, 
the driven stochastic process is a copy of the original process with some modified interactions so that the atypical values of the 
observable in the original process are typical in the driven process. 

In order to construct the driven stochastic generator some basic assumptions have to be considered. Firstly, the value of the
conditioned observable has to satisfy the large deviation principle with a convex rate function and secondly the tilted generator
has to have a spectral gap. In this case, we can construct a driven process via a generalized Doob's h-transform~\cite{CT1,GV}. 
Considering the ensemble-equivalence between micro-canonical and canonical ensembles~\cite{T15,T11,T03,CMT}, equivalence 
of the path ensembles has also recently been studied~\cite{SE}. This reveals that the driven process is a non-conditioned Markov 
process equivalent to the conditioned process with an atypical value of the observable.

It has been shown that the effective generator of a driven Markov process might consist of non-local interactions~\cite{JS}. 
In a recent paper it has been shown that for the time-integrated currents larger than their typical values, non-local interactions 
quite generically lead to a conformally invariant effective dynamics~\cite{KS}. Now the question is 
whether there exist some constraints on the transition rates under which the transitions of the corresponding effective stochastic 
process are local. In other words, the main goal is to find those values of fluctuations for which the dynamics of the system is not 
modified and only the transition rates are rescaled. In~\cite{TJ}, this question is investigated for a family of stochastic Markov 
processes on a one-dimensional lattice with open boundaries. It has also been shown that local interactions appear generically 
in conditioned Zero Range Processes~\cite{HMS}.

In this paper, we firstly aim to answer the above mentioned question for a generic reaction-diffusion model defined on a ring geometry. 
By defining a reaction current as a proper observable, we study its fluctuations in a generalized linear Glauber model with two 
absorbing states. We have looked for the largest eigenvalue of the tilted generator and it turns out that for certain values of the 
conjugated field $s$ the largest eigenvalue of the tilted generator can be calculated analytically using a plane-wave method. 
The left eigenvector associated with this largest eigenvalue has been also obtained. Using these quantities we have been able 
to calculate the current probability distribution in certain regions of the conjugated field $s$. The constants of motion have also been 
calculated.  

In comparison to the previous works~\cite{KJ}-\cite{TJ2} the following remarks have to be mentioned. In~\cite{KJ} the authors 
have studied effective dynamics of generic reaction-diffusion processes defined on a one-dimensional open lattice. They have been 
specifically considered those cases for which the effective interactions are local but site-dependent without referring to any specific 
observable. However, a variant of the zero temperature Glauber model defined on an open lattice with the dynamical rules 
$A\emptyset \to \emptyset \emptyset$ and $A \emptyset \to AA$ where $A$ and $\emptyset$ represent a particle and an 
vacant receptively, is briefly studied as an example. 
In~\cite{MTJ} the authors have consider a variant of the Glauber model (with the dynamical rules as mentioned above) defined on a 
one-dimensional lattice with open boundaries. In this paper, by considering the particle current as a dynamical observable, the dynamical phase 
transitions and the large deviation functions are studied but not the effective dynamics associated with atypical values of the observable. 
The effective dynamics associated with an atypical reaction-diffusion current in generic reaction-diffusion processes, defined on one-dimensional 
open lattices, is studied in~\cite{TJ1}. The authors have specifically looked for the cases where the effective interactions are the same as those 
of the original process (local and site-independent). Different examples are provided but not the Glauber model which will be considered
in the present paper. 
In~\cite{TJ2}, which is a continuation of~\cite{MTJ}, the Glauber model (with the dynamical rules mentioned above) is defined on an open lattice
and the dynamical activity is considered as the observable. The conditional probabilities and the dynamical 
phase transitions are studied in detail. However, the paper is not dealing with the effective interactions.  

This paper is organized as follows: In section~(\ref{sec2}) we start with the mathematical preliminaries. By considering a generic
reaction-diffusion process consisting of classical particles with nearest-neighbor interactions on a one-dimensional lattice with 
periodic boundary conditions, we introduce a reaction-diffusion current as the physical observable in our study. In section~(\ref{sec3}) 
the effective transition rates are obtained for the case where the effective dynamics associated with the original process is local. 
We will introduce a generalized Glauber model in section~(\ref{sec4}). The dynamical behavior of this model is discussed in~(\ref{sec5})  
where we calculate the largest eigenvalue, the right and the left eigenvectors of the associated tilted generator for certain values 
of the conjugated field $s$. In section~(\ref{sec6}) the large deviation function for the current in this model is obtained. In~(\ref{sec7})
we will go back to the effective dynamics of the model. The constants of motion under the effective dynamics are obtained in~(\ref{sec8}). 
In~(\ref{sec9}) we will bring the outlook.
\section{\label{sec2} Dynamical observable and tilted generator}
Consider a classical continuous-time stochastic Markov jump process for which $P(c,t)$ represents the probability of being in 
configuration $c$ at time $t$. The probability distribution vector can be written as 
$|P(t) \rangle\equiv $ $\displaystyle \sum _{c}$$P(c,t)$$|c\rangle$, where $ \{|c\rangle\} $ is a complete basis vector. The master
equation which describes the time evolution of this vector is given by
\begin{equation}
\frac{d}{dt} | P(t)\rangle ={\cal\hat H}| P(t)\rangle \;.
\end{equation}
The stochastic generator ${\cal\hat H}$  has the following matrix elements
\begin{equation}
\langle c|{\cal\hat H} |c^{\prime}\rangle=w_{c^{\prime}\rightarrow c}-\delta_{cc^{\prime}} 
\displaystyle\sum_{c^{\prime\prime}\neq c} w_{c\rightarrow c^{\prime\prime}}
\end{equation}
where $w_{c^{\prime}\rightarrow c}$ is the transition rate from configuration $c^{\prime}$ to $c$ with $w_{c \rightarrow c}=0$.

As we have already mentioned, we are going to investigate the dynamical properties of reaction-diffusion systems of interacting classical 
particles defined on a lattice of length $L$ with periodic boundary conditions by considering the statistics of a time-integrated observable. 
The dynamical observable considered here is the global reaction-diffusion current of particles which will be defined in the following. 
Let us introduce $J$ as a counter which is related to this current during the observation time $t$. Since this quantity is extensive 
with respect to $t$, the value of the total reaction-diffusion current at the end of each trajectory is $j=\frac{J}{t}$. 
In the long-time limit, $j$ might differ from its value in the steady-state $j^{*}$ and the probability of the emergence of atypical 
values $p(j,t)$ becomes exponentially small. Assuming that the large deviation principle is valid then $p(j,t)\sim e^{-tI(j)}$ 
where $I(j)$ is called the Large Deviation Function (LDF).

Let us define $p(c,J,t)$ as the probability of being in the configuration $c$ at time $t$ and the specific value of the counter $J$ is obtained. 
The Laplace transform of this quantity would be $p(c,s,t)= \sum_{J} e^{-sJ}p(c,J,t)$. Hence, by defining the Laplace transform of $p(c,J,t)$
as $p(c,s,t) \equiv \langle c | p_{s}(t)\rangle$, the dynamical partition function 
becomes 
\begin{equation}
\label{key1}
Z(s,t)=\displaystyle \sum _{c} p(c,s,t)= \sum _{c} \langle c | p_{s}(t)\rangle \;.
\end{equation} 
The master equation for the time evolution of $|p_{s}(t)\rangle$ is given by 
 \begin{equation}
 \label{mq}
\frac{d}{dt}|p_{s}(t)\rangle={\cal\hat H}(s)|p_{s}(t)\rangle \;.
\end{equation}
The operator ${\cal\hat H}(s)$ is called non-conservative tilted generator and its matrix elements are defined as 
 \begin{equation}
 \label{key3}
\langle c|{\cal\hat H}(s)|c^{\prime}\rangle=w_{c^{\prime}\rightarrow c}e^{-s\theta_{c^{\prime}\rightarrow c}}-\delta_{cc^{\prime}} 
\displaystyle\sum_{c^{\prime\prime}\neq c} w_{c\rightarrow c^{\prime\prime}}
\end{equation}
in which $\theta_{c^{\prime}\rightarrow c}$ is the increment of the current when the system jumps from $c^{\prime}$ to $c$. 
Using~(\ref{key1})-(\ref{key3}) one finds 
\begin{equation}
Z(s,t)= \sum _{c} p(c,s,t)=\sum_{c} \langle c | e^{t {\cal\hat H}(s) }|p_{s}(0)\rangle \; .
\end{equation}
Assuming that $ {\cal\hat H}(s)$ has a spectral gap, in the long-time limit we find $Z(s,t)\sim e^{t \Lambda^{*}(s)}$ where $\Lambda^{*}(s)$ 
is the largest eigenvalue of ${\cal\hat H}(s)$~\cite{GJLPDW} and that it is the dynamical analogue to the free-energy and predicts the phase 
transitions of the system. At $s=0$, $\Lambda^{*}(s)$ goes to zero and ${\cal\hat H}(s)$ goes back to its stochastic form. On the other hand, at 
$s=0$ we are dealing with the typical value of $j$ in the steady-state while the positive or negative values of $s$ correspond to atypical 
values of current higher or lower than the typical value respectively.

It is worth to mention that the Scaled Cumulant Generating Function (SCGF) of the current is 
the dynamical analogue of the partition function in the canonical Boltzmann-Gibbs approach
\begin{equation}
Z(s,t)=\int  e^{-stj} p(j,t) dj=\langle e^{-sJ} \rangle 
\end{equation}
where the intensive field $s$, which is conjugated to the current, plays a role similar to the inverse temperature $\beta$ 
in the Boltzmann-Gibbs approach. Using $p(j,t)\sim e^{-tI(j)}$ one finds
\begin{equation}
Z(s,t) \sim \int e^{-t(sj+I(j))} dj \sim e^{-t \min \limits_{j} \{ sj+I(j) \} }
\end{equation}
in which $\Lambda^{*}(s)=\max \limits_{j} \{ sj+I(j) \}$. This means that $\Lambda^{*}(s)$ and $I(j)$ are the Legendre-Fenchel transform 
of each other~\cite{T15,LRW}. One should note that the Legendre-Fenchel transform of $I(j)$ yields $\Lambda^{*}(s)$ whether $I(j)$ 
is convex or not; however, If $I(j)$ is nonconvex, then the Legendre-Fenchel transform of $\Lambda^{*}(s)$ does not yield $I(j)$;
rather, it yields the convex envelope of $I(j)$.

The reaction-diffusion systems studied in this paper are modeled using stochastic Markov processes. The stochastic generator of 
these processes consists of nearest-neighbor interactions. The classical particles interact with each other on a one-dimensional lattice 
with periodic boundary conditions according to the following rules:  
\begin{equation}
\label{reactions}
\begin{array}{ccc}
\emptyset \emptyset \overset{w_{21}}{\longrightarrow} \emptyset A \; ,& 
\emptyset \emptyset \overset{w_{31}}{\longrightarrow} A \emptyset  \; ,& 
\emptyset \emptyset \overset{w_{41}}{\longrightarrow} A A \; , \\
\emptyset A \overset{w_{12}}{\longrightarrow} \emptyset \emptyset \; ,& 
\emptyset A \overset{w_{32}}{\longrightarrow} A \emptyset  \; ,& 
\emptyset A \overset{w_{42}}{\longrightarrow} A A \; , \\
A \emptyset \overset{w_{13}}{\longrightarrow} \emptyset \emptyset \; ,& 
A \emptyset \overset{w_{23}}{\longrightarrow}  \emptyset  A \; ,& 
A \emptyset \overset{w_{43}}{\longrightarrow} A A \; , \\
AA \overset{w_{14}}{\longrightarrow} \emptyset \emptyset \; ,& 
AA \overset{w_{24}}{\longrightarrow}  \emptyset A \; ,& 
AA \overset{w_{34}}{\longrightarrow} A \emptyset 
\end{array}
\end{equation}
in which $w_{ij}$s' are the rates at which the reactions occur. For our later convenience we also define $w_{ii}=-\Sigma_{j\neq i} w_{ji}$. 

Following~\cite{TJ}, the increments of the reaction-diffusion current, which will be considered as a proper dynamical observable throughout 
this paper, are defined as follows
\begin{equation}
\label{rules}
\begin{array}{ll}
\theta_{\emptyset \emptyset   \longrightarrow \emptyset  A } = +1 , \ \ \ \ \ \ \theta_{A A  \longrightarrow \emptyset A } = +1,\\
\theta_{\emptyset \emptyset  \longrightarrow  A  \emptyset } = -1 ,  \ \ \ \ \ \ \theta_{A A \longrightarrow A \emptyset }= -1,\\
\theta_{\emptyset A  \longrightarrow  \emptyset  \emptyset } = -1 ,  \ \ \ \ \ \ \theta_{A \emptyset  \longrightarrow  \emptyset  \emptyset  } = +1,\\
\theta_{\emptyset A  \longrightarrow A \emptyset  } = -1 ,               \ \ \ \ \ \theta_{A  \emptyset \longrightarrow \emptyset  A } = +1,\\
\theta_{\emptyset  A   \longrightarrow   A A } = -1 ,                             \ \ \ \ \theta_{A  \emptyset  \longrightarrow A A } = +1.\\
\end{array}
\end{equation}

\section{\label{sec3} Effective interactions}
As we have mentioned before, there exists a Time Translational Invariant regime (TTI), during which the mean value of our observable 
over the biased ensemble of trajectories is identical to the corresponding value in the steady-state of a non-conditioned driven (effective) 
process. In this section, we focus on constructing the effective Hamiltonian for reaction-diffusion systems explained in previous section. 
Let us call $ \langle\tilde{\Lambda}^{*}(s)|$ and $ |{\Lambda}^{*}(s)\rangle$ as the left and right eigenvectors of ${\cal\hat H}(s)$ and 
$\Lambda^{*}(s)$ as its largest eigenvalue corresponding to those vectors. According to the Doob's h-transform~\cite{CT1}, we can write the 
effective Hamiltonian as
\begin{equation} \label{heff}
{\cal\hat H}_{eff}(s)=\hat{V} {\cal\hat H}(s) \hat{V}^{-1}-\Lambda^{*}(s)
\end{equation}
in which $\hat{V}$ is a diagonal matrix with $\langle c|\hat{V}|c\rangle =\langle\tilde{\Lambda}^{*}(s)|c\rangle$. Now the effective rates can 
be deduced from~(\ref{heff})
\begin{equation} \label{weff}
w^{eff}_{c \to c^{\prime}}(s)=w_{c\to c^{\prime}} e^{-s \theta_{c \to c'}} \frac{\langle\tilde{\Lambda}^{*}(s)|c\rangle}
{\langle\tilde{\Lambda}^{*}(s)|c^{\prime}\rangle} \; .
\end{equation}
It is clear that an effective process associated with a process with two-site interactions do not necessarily contain the same local interactions. 
In other words, there might be long-range interactions in the driven process. In this paper, we are looking for those values 
of $s$ for which the effective interactions are local and the transition rates are simply rescaled values of those of the original process. 
One of the most straightforward approaches is to consider $\langle S | \hat{\cal H} (s^\ast) = 0$ where $ \langle S|=(1\ 1\ 1\ ....1)$ ~\cite{TJ}. 
This equation determines the value(s) of the conjugated field $s^\ast$ for which the effective interactions are local given that the interactions 
in the original process are local. This way of determining $s^\ast$ is not unique i.e. there might be other approaches using which one can find 
those values of $s^\ast$ for which the locality of effective interactions is fulfilled. We will come back to this point in section~\ref{rev}. 
It turns out that for the models considered here $s^{*}$ is
\begin{equation}
\label{s^{*}}
s^{*}=\ln[\frac{-2w_{34}+w_{12}+w_{32}+w_{42}}{-2w_{24}+w_{13}+w_{23}+w_{43}}].
\end{equation}
provided that the fraction in the right hand side of the above equation is positive. The corresponding largest eigenvalue is
\begin{equation} \label{cons}
\Lambda ^{*}(s^{*})=L(w_{11}+w_{21}e^{s^{*}}+w_{31}e^{-s^{*}}+w_{41})\;.
\end{equation}
In what follows, we investigate a generalization of the zero-temperature Glauber model on a ring geometry which consists of non-zero
reaction rates $w_{12}$, $w_{24}$, $w_{31}$ and $w_{43}$. As we mentioned in section~\ref{sec1}, the specific variant of the 
Glauber model which has already been studied in~\cite{KJ,MTJ,TJ2} consist of only $w_{13}$ and $w_{43}$. 

\section{\label{sec4} A generalized Glauber model}
The Linear Glauber Model (LGM) is a spin system with three-site interactions in $d$-dimension with the following one-spin-flip rate~\cite{O}
\begin{equation}
\label{vm rates}
w^{LGM}_{i}(\sigma)=\frac{\alpha}{2}(1-\frac{\gamma}{2d}\sigma_{i}\sum_{\delta}\sigma_{i+\delta}) \;.
\end {equation}
in which the sum runs over the nearest neighbors with $\sigma_{i}=\pm1$. In~(\ref{vm rates}) $\alpha$ defines the time 
scale, $\gamma= \tanh(2\beta E)$ with $E$ being the energy to create a domain wall in a domain of uniform magnetization and $\beta$ is
the inverse of temperature. For $d=1$ we have
\begin{equation}
w^{LGM}_{i}(\sigma)=\frac{\alpha}{2}(1-\frac{\gamma}{2}\sigma_{i}(\sigma_{i-1}+\sigma_{i+1})) \;.
\end{equation}
At zero temperature this model can then be mapped onto a two-site reaction-diffusion model by considering an up-spin as a particle and a down-spin 
as a vacancy with the following dynamical rules that all take place with the same rate~\cite{SHZ} 
$$
\begin{array}{ccc}
\emptyset A  {\longrightarrow} \emptyset \emptyset \; ,& 
\emptyset A  {\longrightarrow} A A \; , \\
A \emptyset  {\longrightarrow} \emptyset \emptyset \; ,& 
A \emptyset  {\longrightarrow} A A \; . \\
\end{array}
$$
As a generalization to this model we consider a model with the following dynamical rules
$$
\begin{array}{ccc}
\emptyset A \overset{w_{12}}{\longrightarrow} \emptyset \emptyset \; ,& 
\emptyset A \overset{w_{42}}{\longrightarrow} A A \; , \\
A \emptyset \overset{w_{13}}{\longrightarrow} \emptyset \emptyset \; ,& 
A \emptyset \overset{w_{43}}{\longrightarrow} A A \; , \\
\end{array}
$$
where the reaction rates are not necessarily equal. In the following sections we consider this generalized Glauber model on a one-dimensional 
lattice with periodic boundary conditions and the reaction current, whose increments are defined by~(\ref{rules}), as the observable. This current 
can be explained as follows: Let us consider a single patch of consecutive particles on the ring, for instance a configuration similar to 
$\cdots\emptyset\emptyset AAAAAA\emptyset\emptyset\cdots$. 
According to the dynamical rules this patch can only be increased or decreased in size from its both ends. Any creation or annihilation of 
particle that takes place at the head of this patch (let us say the right end) will increase the current. In contrast, any creation or annihilation 
of particle that takes place at the tail of this patch (let us say the left end) will decrease the current. The dynamical
rules does not allow any kind of reaction in the middle of the patch; therefore, the total current can be interpreted as a net activity of each patch 
which will be decreased by the reactions at the tail and increased by the reactions at the head of each patch. In what follows we will use the 
term {\em current} for this observable. 

\section{\label{sec5} Dynamical behavior of the model}
\subsection{Mean-field approximation}
\begin{figure*}
\centering
    \includegraphics[width=.3\linewidth]{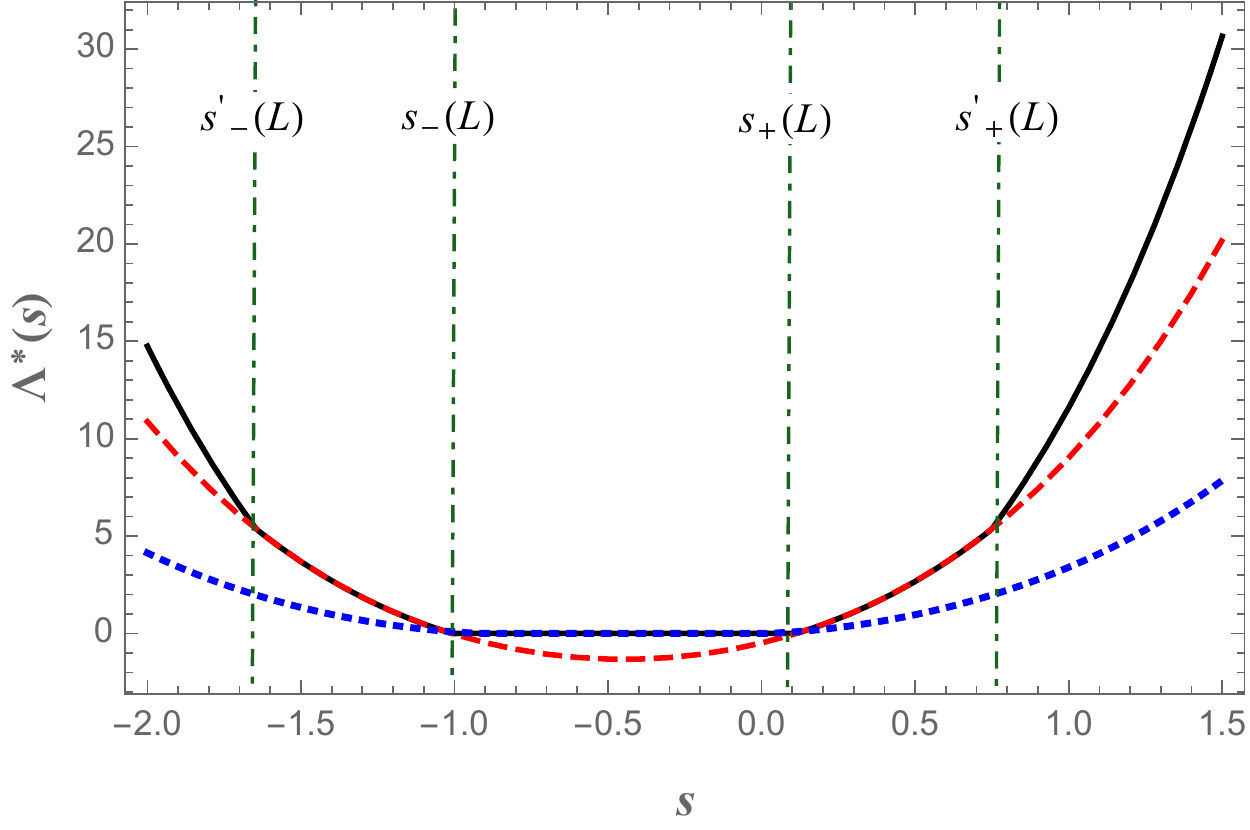}
    \includegraphics[width=.312\linewidth]{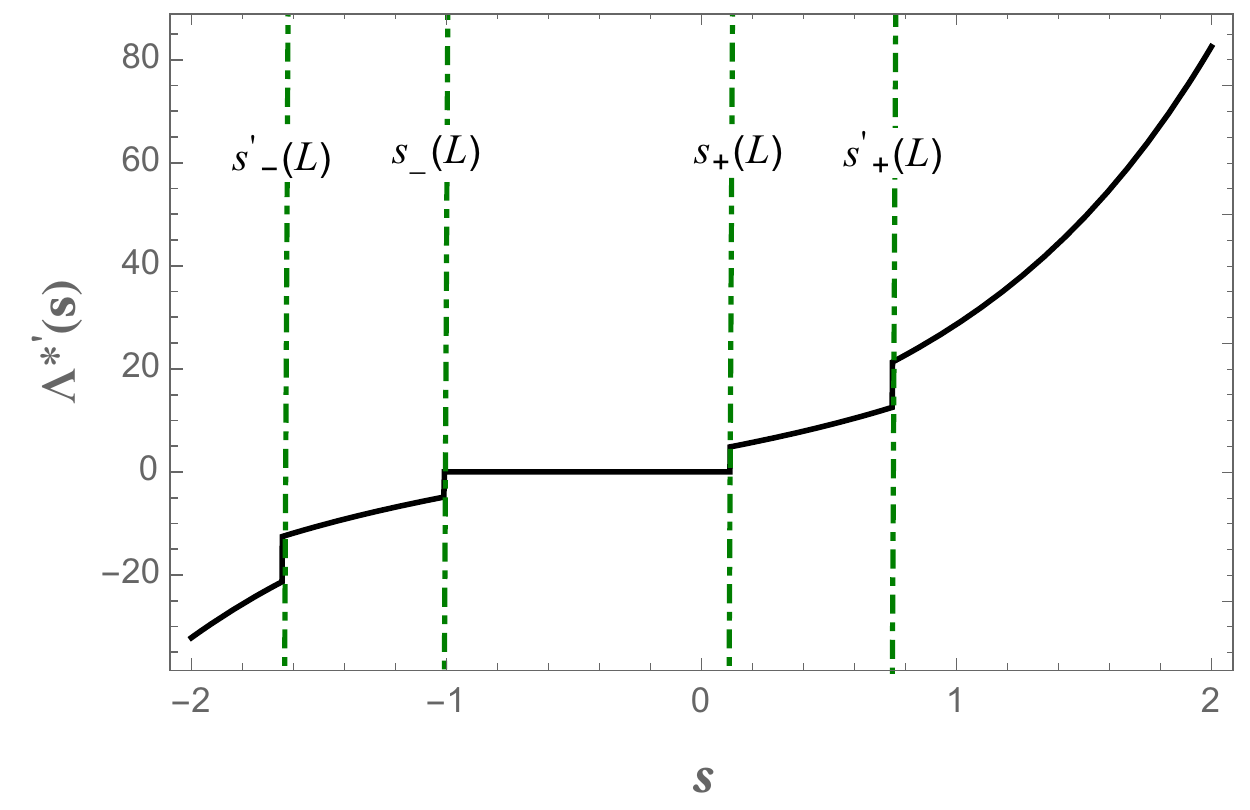}
    \includegraphics[width=.3\linewidth]{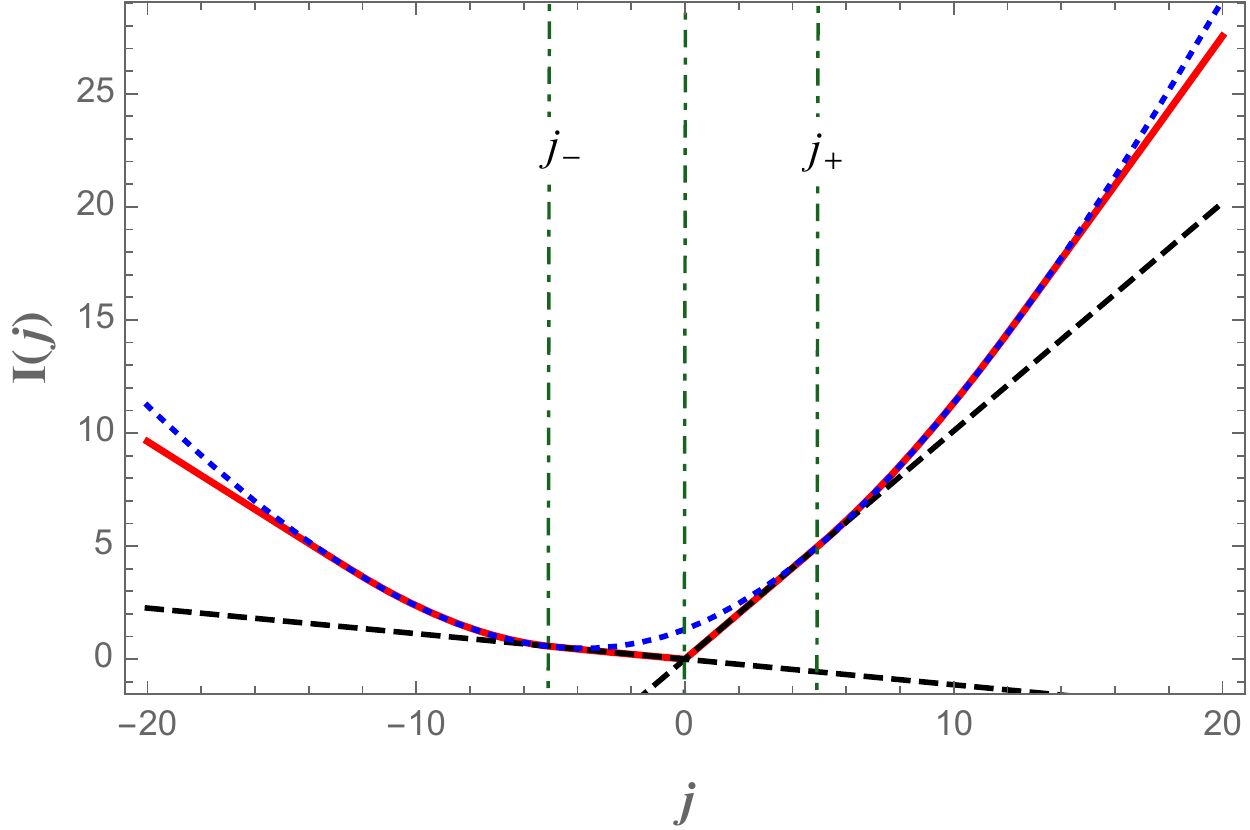}
    \caption{\label{fig1} (Color online) For a chain of length $L=10$ with $w_{12}=1$, $w_{42}=2$, $w_{13}=4$ and $w_{43}=3$  
    the leftmost figure shows the largest eigenvalue of ${\cal\hat H}(s)$ as a function of $s$. The solid line is the numerical 
    solution, the dashed line is the plane-wave solution~(\ref{lambda}) and the dotted line is the mean-field approximation. The validity 
    domain of the plane-wave solution is shown by the vertical dashed lines. The middle figure is the derivative of the largest 
    eigenvalue $\Lambda^{*'}(s)$ and the phase transition points obtained from the exact numerical results. The rightmost figure 
    illustrates the LDF as a function of $j$. The solid line is obtained numerically while the 
    dashed lines are~(\ref{j1}) and~(\ref{j2}). The dotted line is the Legendre-Fenchel transform of~(\ref{lambda}) which, as can be seen, is 
    valid for certain values of $j$. For more information see inside the text.}
\end{figure*}

As it was mentioned, $\Lambda^{*}(s)$ plays the role of the dynamical free-energy of the system and its singularities determine the dynamical 
phase transitions of the system. In order to have a preliminary insight, let us find $\Lambda^{*}(s)$ in the mean-field approximation. Using the 
Doi-Peliti formalism~\cite{DP}, ${\cal\hat H}(s)$ can be rewritten in terms of the bosonic creation and annihilation operators $a_{i}$ and $a_{i}^{\dagger}$ 
as follows
\begin{eqnarray}
\label {BSH}
{\cal\hat H}(s)&=&\sum_{i=1}^{L}w_{12}(1-n_{i})a_{i+1}(e^{-s}-a_{i+1}^{\dagger}) \\
&+&w_{42}n_{i+1}(e^{-s}a_{i}^{\dagger}-1)\nonumber  \\
&+&w_{13}(1-n_{i+1})a_{i}(e^{s}-a_{i}^{\dagger}) \nonumber \\
&+&w_{43}n_{i}(e^{s}{a_{i+1}}^\dagger-1) \nonumber 
\end{eqnarray}
where $n_{i}=a_{i} a_{i}^{\dagger}$ is the number operator. Replacing the operators by constant fields as $a_{i} \to \langle a_{i}\rangle\equiv \phi$ 
and $a_{i}^{\dagger} \to \langle a_{i}^{\dagger}\rangle\equiv\overline{\phi}$ in~(\ref{BSH}) one finds
\begin{equation}\label{HF}
{\cal\hat H}(\phi,\overline{\phi},s)=(1-\phi \overline{\phi} )\phi(D_{a}-\overline{\phi}g_{a})+\phi \overline{\phi} (D_{c} \overline{\phi}-g_{c})
\end{equation}
where $g_{a}=w_{12}+w_{13}$, $g_{c}=w_{42}+w_{43}$, $ D_{a} =w_{12} e^{-s} + w_{13} e^{s} $  and $ D_{c} =w_{42} e^{-s} + w_{43} e^{s} $. 
The largest value of~(\ref{HF}) can be obtained by calculating its first derivative with respect to $\phi$ and $\overline{\phi}$ 
$$
    \left\{
\begin{array}{l}
(D_{a}-\overline{\phi} g_{a})(1-2\phi \overline{\phi})+(\overline{\phi} D_{c}-g_{c})\overline{\phi}=0\; ,\\ \\
\phi (\overline{\phi} D_{c}-g_{a})-\phi^{2}(D_{a}-2\overline{\phi} g_{a})+\phi(\overline{\phi} D_{c}-g_{c})=0\;.\\
\end{array}
  \right.
$$
Substituting the solution of these equations into~(\ref{HF}) gives $\Lambda^{*}(s)$. It turns out that $\Lambda^{*}(s)$ is a two-criterion function 
as $s$ varies
$$
    \left\{
    \begin{array}{ll}
    \Lambda^{*}(s)=0 \quad \mbox{for} \qquad s_{-}(\infty) \le s \le s_{+}(\infty) \; , \\ \\
    \Lambda^{*}(s)\neq0 \quad \mbox{for} \qquad s < s_{-}(\infty) \quad \mbox{and} \quad s > s_{+}(\infty) \\
    \end{array}
    \right.
$$
in which $s_{\pm}(\infty)$ are given by
\begin{equation} 
\begin{array}{lll}
s_{\pm}(\infty)=\frac{1}{2} [\ln (\frac{1}{8 w_{13} w_{43}} (\delta^\prime \pm \sqrt{(\delta^\prime)^{2}-(\gamma^\prime) ^{2}})] \\
\end{array}
\end{equation} 
where $\delta^\prime=g^{2}-4(w_{12}\ w_{43}+w_{42}\ w_{13})$ and $\gamma^\prime=8\sqrt{w_{12}\ w_{42}\ w_{13}\ w_{43}}$. 
It is easy to check that two first-order phase transitions take place at $s_{\pm}(\infty)$ since the first derivative of $ \Lambda^{*}(s)$ changes 
discontinuously at these points (see FIG.\ref{fig1}). 
\subsection{\label{pw} Plane-wave method}
Assuming that the model contains classical particles which evolve in time under the dynamical rules of the generalized Glauber model, in 
this section, we aim to diagonalize the tilted generator associated with the current defined in~(\ref{rules}) and find its largest eigenvalue 
corresponding to each value of $s$. 

It turns out that the configuration space of our model has a hierarchical structure with two absorbing states. 
This means that the configuration space is not ergodic. The eigenvectors 
of $\hat{\cal H}(s)$ corresponding to these states will be denoted by $\vert L \rangle$ and $\vert 0 \rangle$ respectively so that 
$\hat{\cal H}(s)  \vert 0\rangle= \hat{\cal H}(s)\vert L \rangle =0 $. The largest eigenvalue of the tilted generator is 
zero in a certain values of $s$ whose domain of validity will be clear later. The next sub-space of the configuration space consists of the states 
containing $k$ consecutive particles (or a single patch of length $k$) on the ring. The dynamical rules do not let a single patch of $k$ 
consecutive particles to be divided into smaller patches. Instead, a single patch can either merge into an empty lattice or grow until the lattice is fully 
occupied by the particles. Therefore, if one starts from a state of $k$ consecutive particles, which will be denoted by $\vert k \rangle $ with $1 \le k \le L-1$, 
it evolves under the dynamical rules in this $(L-1)$-dimensional sub-space of the configuration space until it traps into one of the absorbing states. 
There is also a sub-space of the configuration space which consists of the configurations with two blocks of consecutive particles which 
are at least one particle apart. A member of this sub-space, which will be denoted by $\vert k, k^{'} \rangle$ with $1 \le k,k' \le L-3$ and $k+k' \le L-2$, 
can be schematically shown as
$$\cdots\emptyset \emptyset \emptyset \overbrace{A A \cdots A A }^{k} \emptyset \emptyset \cdots \emptyset \overbrace 
{A A \cdots A A }^{k^{'}} \emptyset \emptyset \cdots  \; .$$
A member of the sub-space $\{ \vert k,k' \rangle\}$ can evolve in time until it drops into the sub-space $\{ \vert k \rangle \}$. The reader can easily see that the 
last sub-space of the configuration space, considering this type of grouping of states, has only one member and it is the state of 
$\cdots \emptyset A \emptyset A \emptyset A \cdots$. 

\begin{figure}
    \centering
    \includegraphics[width=80mm]{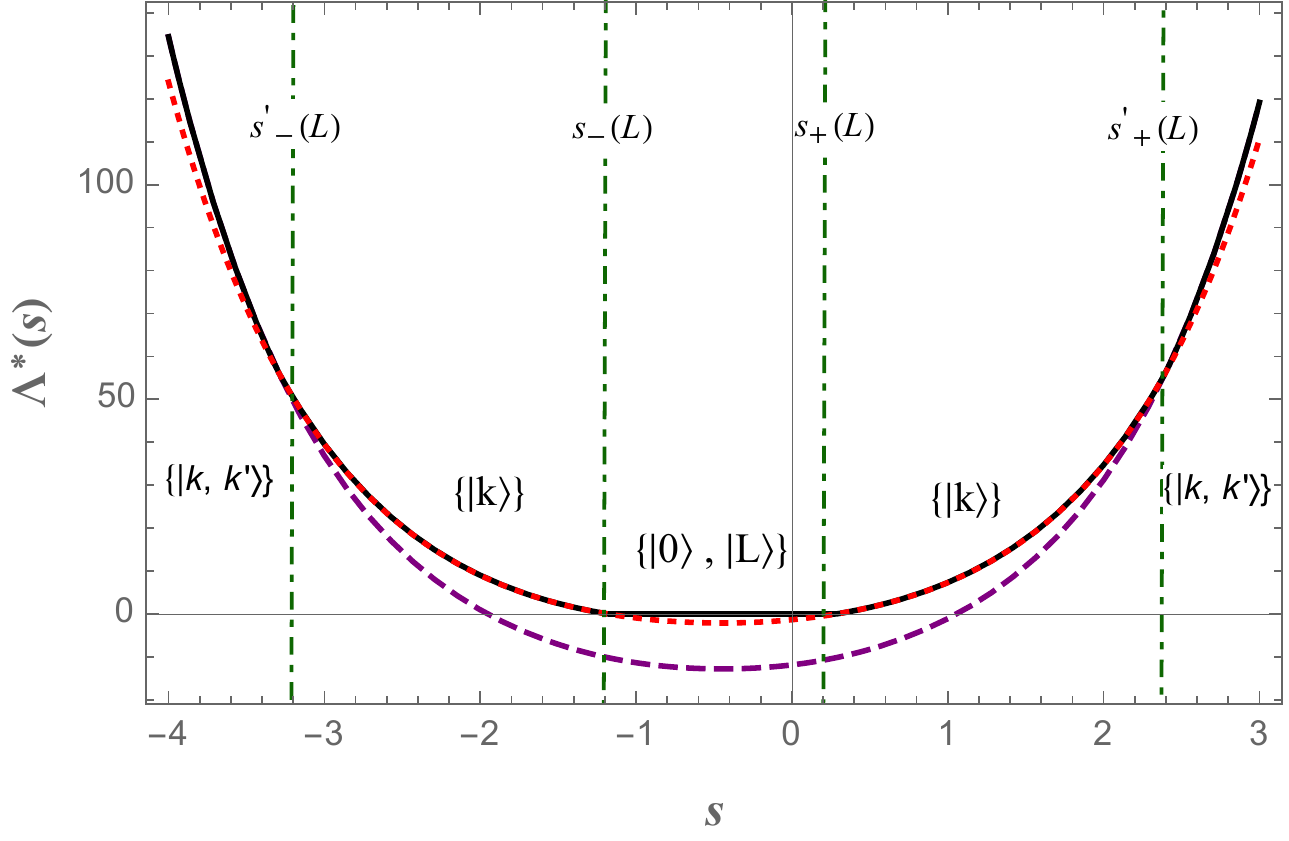}
    \caption{\label{fig4} (Color online) The largest eigenvalue of ${\cal\hat H}(s)$ as a function of $s$ for a chain of length $L=6$  
    with $w_{12}=1$, $w_{42}=2$, $w_{13}=4$ and $w_{43}=3 $. The solid line comes from numerical diagonalization of ${\cal\hat H}(s)$, the dotted 
    line is the plane-wave solution in the sub-space spanned by $\{\vert k \rangle\}$ and the dashed line is the numerical solution in the sub-space 
    spanned by $\{\vert k, k^{'}\rangle\}$. The vertical dashed lines separate these regions.}
\end{figure}

Let us first diagonalize $\hat{\cal H}(s)$ in $\{ \vert k \rangle\}$ with $1\le k\le L-1$. It can easily be seen that 
\begin{equation}
\begin{array}{lll}
\label{bethe}
\hat{\cal H}(s)|1 \rangle =D_{c} |2 \rangle -g|1  \rangle \; ,\\ 
\hat{\cal H}(s)|k \rangle =D_{a} |k-1 \rangle +D_{c} |k+1 \rangle -g|k  \rangle \; ,\\ 
\hat{\cal H}(s)|L-1 \rangle =D_{a} |L-2 \rangle -g|L-1  \rangle \; ,
\end{array}
\end{equation}
in which $2\le k \le L-2$ and $g=g_a+g_c$. The right eigenvector of $\hat{\cal H}(s)$ in this sub-space can be written as
\begin{equation}
 | \Lambda(s) \rangle = \sum_{k=1}^{L-1} C_{k}(s) |k \rangle\;.
\end{equation}
Considering $\hat{\cal H}(s) \vert  \Lambda (s)  \rangle =  \Lambda (s) \vert  \Lambda (s)  \rangle $ we find the equations governing 
$C_{k}(s)$'s as follows
\begin{eqnarray}
\label{recur}
&& \Lambda (s) C_{1}(s) = D_{a} C_{2}(s) -g C_{1}(s)\; , \nonumber \\
&& \Lambda (s) C_{k}(s) = D_{c} C_{k-1}(s) +D_{a} C_{k+1}(s) -g C_{k}(s)\; , \nonumber  \\
&& \Lambda (s) C_{L-1}(s) = D_{c} C_{L-2}(s) -g C_{L-1}(s) \; .
\end{eqnarray}
According to the plane-wave method we assume that $C_{k}(s)$'s have the following form~\cite{SHZ}
\begin{equation}
C_{k}(s)= \alpha z_1^{k}+\beta z_2^{k} \; .
\end{equation}
After some straightforward calculations (see~\cite{SZ} for instance) we finally find the largest eigenvalues of $\hat{\cal H}(s)$ in this sub-space 
\begin{equation}
 \label{lambda}
\Lambda^{\ast}(s)=2\sqrt{D_{a} D_{c}}\cos(\frac{\pi}{L})-g\;  .
\end{equation}
It is clear that~(\ref{lambda}), which comes from diagonalizing $\hat{\cal H}(s)$ in the above mentioned sub-space
covers only a certain region of $s$. This region is shown in~FIG.\ref{fig1} which starts from $s_{-}(L)$ to $s^{'}_{-}(L)$ 
(whose value comes from diagonalizing $\hat{\cal H}(s)$  in the sub-space $\{\vert k, k^{'}\rangle\}$) for which $j<0$ and from $s_{+}(L)$ to 
$s^{'}_{+}(L)$ (whose value comes from diagonalizing $\hat{\cal H}(s)$ in the sub-space $\{\vert k, k^{'}\rangle\}$) for which $j>0$. Solving
$\Lambda^{\ast}(s)=0$ one can find $s_{\pm}(L)$. Defining 
$$\delta=g^{2}-2(\cos(\frac{2\pi}{L})+1)(w_{12}\ w_{43}+w_{42}\ w_{13})$$ 
and 
$$\gamma=8\sqrt{w_{12}\ w_{42}\ w_{13}\ w_{43}}\cos^{2}(\frac{\pi}{L})$$ 
it turns out that
\begin{equation} 
\label{spm}
\begin{array}{lll}
s_{\pm}(L)=\frac{1}{2}\ln (\frac{\sec^{2}(\frac{\pi}{L})}{8 w_{13} w_{43}}(\delta\pm \sqrt{\delta^{2}-\gamma^{2}})) \; .
 \end{array}
\end{equation} 
It can be seen that as $L \to \infty$ the exact results approach to those obtained from the mean-field approximation. 
Finding $s^{'}_{+}(L)$ and $s^{'}_{-}(L)$ analytically is more complicated. They can be found by diagonalizing $\hat{\cal H}(s)$ in a sub-space
spanned by $\{ \vert k , k' \rangle\}$. In FIG.\ref{fig4}, numerically obtained $\Lambda^{*}(s)$ has been plotted. 
Diagonalizing $\hat{\cal H}(s)$ in $\{ \vert k , k' \rangle\}$ gives the largest eigenvalue of the tilted generator only between 
$s^{'}_{\pm}(L)$ and some $s^{''}_{\pm}(L)$ (not plotted here) which has to be obtained separately. Note that the largest 
eigenvalue of the tilted generator in the region $s_{-}(L) \le s \le s_{+}(L)$ is zero. 

As we mentioned the discontinuities of the first derivative of $\Lambda^{*}(s)$ are associated with the first-order dynamical phase transitions. 
In FIG.\ref{fig1} the first derivative of $\Lambda^{*}(s)$ is plotted. It can be seen $\Lambda^{*'}(s)$ is discontinuous at the transition points. 
Between $s_{-}(L)$ and $s_{+}(L)$ the average current is zero. At $s_{+}(L)$ ($s_{-}(L)$) the average current jumps from zero to some finite positive 
(finite negative) value. As a matter of fact, a hierarchy of first-order dynamical phase transitions occur whenever one goes from one sub-space 
to another sub-space.
\section{\label{sec6} Large deviation function}
In this section we calculate the large deviation function $I(j)$ in the regions where $\Lambda^{*}(s)$ has been obtained from analytical 
diagonalization of tilted generator. The large deviation function has to be calculated from the Legendre-Fenchel transform of $\Lambda^{*}(s)$ 
given by~(\ref{lambda})~\cite{T03}. We note that the first derivative of $\Lambda^{*}(s)$ is not continuous at $s_{-}(L)$ and $s_{+}(L)$. Since 
the slope of the line which connects these two points is equal to zero; therefore, the minimum of the large deviation function is at $j=0$. On 
the other hand, because of that discontinuity, the large deviation function will have two linear parts. For $0 \le j \le j_{+}$ the large deviation
function is given by 
\begin{equation}
\label{j1}
I(j)=s_{+}(L)\, j
\end{equation}
while for $j_{-} \le j \le 0$ it is given by 
\begin{equation}
\label{j2}
I(j)=s_{-}(L)\, j 
\end{equation}
in which $s_{\pm}(L)$ comes from~(\ref{spm}) and that 
$$
 j_{+}= \lim_{s \searrow s_{+}(L)} \frac{d}{ds} \Lambda^{*}(s) \;\; \mbox{and} \;\;
 j_{-}= \lim_{s \nearrow s_{-}(L)} \frac{d}{ds} \Lambda^{*}(s)  \; .
$$
For $j_{+} \le j \le j_{+}^{'}$ ($j_{-}^{'} \le j \le j_{-}$) the large deviation has to be calculated from the Legendre-Fenchel transform 
of~(\ref{lambda}) in $s_{+}(L) \le s \le s_{+}^{'}(L)$ ($s_{-}^{'}(L) \le s \le s_{-}(L)$) for which we have
$$
 j_{+}^{'}= \lim_{s \nearrow s_{+}^{'}(L)} \frac{d}{ds} \Lambda^{*}(s) \;\; \mbox{and} \;\;
 j_{-}^{'}= \lim_{s \searrow s_{-}^{'}(L)} \frac{d}{ds} \Lambda^{*}(s)  \; .
$$
Now, similar to the previous case, since the first derivative of largest eigenvalue of the tilted generator is not continuous at $ s_{\pm}^{'}(L)$ 
(because it comes from different sub-spaces of the configuration space), the large deviation function will have two linear parts 
between $j'_{+} \le j \le j''_{+}$ and $j''_{-} \le j \le j'_{-}$. This scenario will be repeated as $s$ increases (decreases) toward $+\infty$ ($-\infty$).
\section{\label{sec7} effective dynamics revisited \label{rev}}
As we mentioned, the point at which the effective dynamics of the model is local and that the effective transition rates are rescaled 
values of those of the original process can be obtained from~(\ref{s^{*}}) 
\begin{equation}\label {sstar}
 s^{*}=\ln[\frac{w_{12}+w_{42}}{w_{13}+w_{43}}]
 \end{equation}
with $\Lambda^{*}(s^{*})=0$. Note that for the case $w_{12}=w_{13}$ and $w_{42}=w_{43}$, $s^{\ast}$ 
disappears and a stochastic process with the above mentioned properties does not exist. However, $s^{\ast}$ is generally non-zero in our model.

\begin{figure}[t]
    \centering
    \includegraphics[width=70mm]{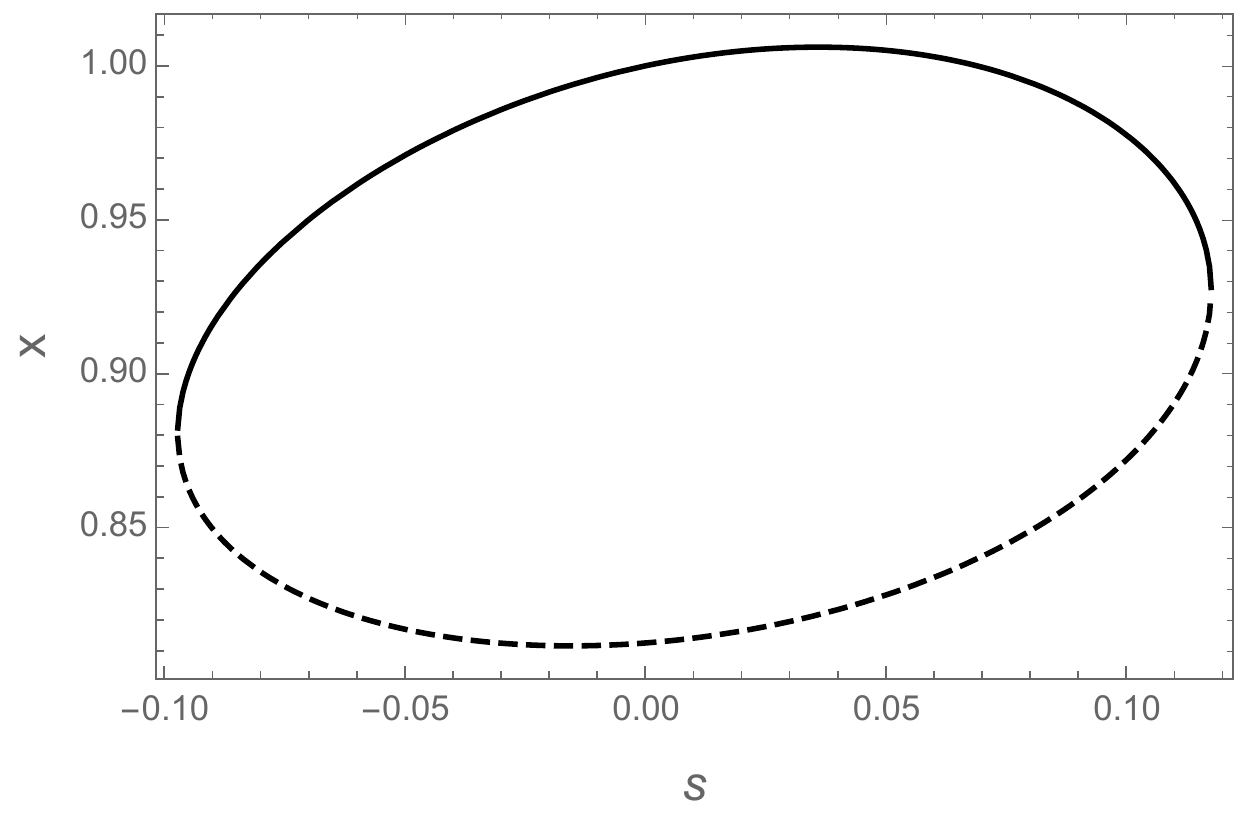}
    \caption{\label{fig3} $x$ as a function of $s$ for $w_{12}=2.5$, $w_{42}=5$, $w_{13}=4$, $w_{43}=3$. 
    $x_{+}$ is plotted by a solid line while $x_{-}$ by a dashed line. The domain of validity is $s_{-}(\infty) \le s \le s_{+}(\infty)$ ($-0.097 \le s \le 0.118$). 
    Note that $s^{*}=0.069$.}
\end{figure}

Let us now consider the following left eigenvalue of ${\cal\hat H}(s)$ 
\begin{equation}
\langle X \vert \equiv (1\;\; x)^{\otimes L} 
\end{equation}
with zero eigenvalue so that $\langle X \vert  {\cal\hat H}(s)=0$. After some calculations one finds that $x$ can take two values $x_{\pm}(s)$ given by
\begin{equation}
x_\pm(s)=\frac{g\pm \sqrt{g^2-4 D_a D_c}}{2 D_c} \; .
\end{equation}
These solutions are valid as long as $s$ takes real values. In FIG.~\ref{fig3} we have plotted $x_{\pm}(s)$ as function of $s$ for certain values of the 
parameters. The domain of validity of the obtained solutions is $s_{-}(\infty) \le s \le s_{+}(\infty)$. These values of $x$ introduce two effective dynamics 
for each given $s$ in the mentioned domain.  

Using~(\ref{heff}) one can see that in the effective process the creation rates of the biased process are multiplied by $x$ and the annihilation rates are 
multiplied by $1/x$. As a result, the dynamical rules of the effective dynamics will be
$$
\begin{array}{ccc}
\emptyset A \overset{w_{12}e^{-s}x^{-1}}{\xrightarrow{\hspace*{1.5cm}}} \emptyset \emptyset \; ,& 
\emptyset A \overset{w_{42}e^{-s}x}{\xrightarrow{\hspace*{1.5cm}}} A A \; ,  \\
A \emptyset \overset{w_{13}e^{s}x^{-1}}{\xrightarrow{\hspace*{1.5cm}}} \emptyset \emptyset \; ,& 
A \emptyset \overset{w_{43}e^{s}x}{\xrightarrow{\hspace*{1.5cm}}} A A \; . \\
\end{array}
$$
This means that the interactions are still local and the original transition rates are only rescaled. 

At $s=0$ (note that $s=0$ is always in $[s_{-}(\infty) ,s_{+}(\infty)]$) we find out that one of the $x$'s is equal to unity which reflects the fact that the 
original generator of the process is stochastic (see FIG.~\ref{fig3}). Assuming that at $s=0$ we have $D_a<D_c$ then $x_{+}(0)=1$ and that $x_{-}(0)=D_a/D_c$ for 
which we find the constant of motion given by 
$$(1\;\; x_{-}(0))^{\otimes L}  \vert P(t)\rangle\; .$$
If $p_n(t)$ is defined as the probability of existing exactly $n$ particles ($0 \le n\le L$) on the lattice at the time $t$ then the constant of motion is given by
$$ \sum_{n=0}^{L} (\frac{D_a}{D_c})^n p_n(t) \; .$$

\section{\label{sec8} Constants of motion in the effective dynamics}

In order to find the constants of motion for the effective process where the average current is zero, the left eigenvectors of the effective stochastic 
Hamiltonian with the zero eigenvalue have to be calculated. Similar to the previous section, we consider a left eigenvector for the effective process 
as follows 
 \begin{equation}\label{xeff}
 \langle X_{eff}| \hat{ \cal {H}}_{eff} (s)=0\;
 \end{equation}
in which $ \langle X_{eff}| \equiv (1\;\; y)^{\otimes L} $. It turns out that $y$ can take three values of $1$, $D_{a}/D_{c}$ and $D_{c}/D_{a}$. 
Similar to the previous section, let us write the probability vector at any time as
\begin{equation}
\label{vec}
\vert P_{eff}(t)\rangle = \sum_{n=0}^{L} {p_{n}(t)} \vert p_{n}\rangle
\end{equation}
where $p_{n}(t)$ has the above mentioned definition and that $\vert p_{n}\rangle$ is a basis vector. The 
time evolution of~(\ref{vec}) is governed by the master equation for the effective process
\begin{equation}
\label{mqe}
\frac{d}{dt} \vert P_{eff}(t)\rangle = \hat{ \cal {H}}_{eff} (s) \vert P_{eff}(t)\rangle \; .
\end{equation}
The constants of motion are now given by $\langle X_{eff}| P_{eff}(t)\rangle $. Using the above results we find two constants of motion 
\begin{equation}
\sum_{n=0}^{L} {(\frac{D_a}{D_c})^{n} p_{n}(t)} \;\; \mbox{and} \;\; \sum_{n=0}^{L} {(\frac{D_c}{D_a})^{n} p_{n}(t)}\;. 
\end{equation}
If $D_a=D_c$ the constants of motion are equal to unity which expresses the conservation of probability. 

\section{\label{sec9} Summery and Outlook}
In this paper, the fluctuations of a current in a stochastic Markovian system of classical particles with two-site interactions under periodic boundary 
conditions has been studied. An effective model has been built by modifying a stochastic Markov process conditioned on an atypical value of the 
current. The resulting process might have non-local interactions. We have obtained some constraints under which the interactions in the driven 
process is local. In other words, we have found at least one specific effective stochastic process that obeys exactly the same dynamical rules as 
the original process but with rescaled reaction rates. To verify our approach, a generalized Glauber model has been considered and its effective 
stochastic generator has been obtained under certain conditions. We have investigate the dynamical phase transitions in this model. 
Moreover, constants of motion in the effective processes are calculated. 

The problem of finding the effective dynamics associated with a rare event is an ongoing and active field of research. The effective interactions 
between classical particles hopping on a homogeneous ring has already been studied~\cite{PSS}. It would be interesting to consider the models 
consisting of diffusive particles on a ring in the presence of both particle and/or lattice-site defects and investigate the effective interactions between 
the particles in these cases. 


\end{document}